\documentclass[pdflatex,sn-mathphys-num]{sn-jnl}

\usepackage[version=4]{mhchem}
\usepackage{caption}
\usepackage{geometry}
\usepackage{enumitem}
\usepackage{lineno}
\usepackage{numprint}
\npthousandsep{,}
\usepackage{xr}
\usepackage{soul, color, xcolor}
\soulregister{\cite}7
\soulregister{\ref}7
\soulregister{\ce}7
\soulregister{\numprint}7
\usepackage{hyperref}

\usepackage{graphicx}%
\usepackage{multirow}%
\usepackage{amsmath,amssymb,amsfonts}%
\usepackage{amsthm}%
\usepackage{mathrsfs}%
\usepackage[title]{appendix}%
\usepackage{xcolor}%
\usepackage{textcomp}%
\usepackage{manyfoot}%
\usepackage{booktabs}%
\usepackage{algorithm}%
\usepackage{algorithmicx}%
\usepackage{algpseudocode}%
\usepackage{listings}%
\theoremstyle{thmstyleone}%
\theoremstyle{thmstyletwo}%
\theoremstyle{thmstylethree}%
\modulolinenumbers[1]

\newcommand{\onlinecite}[1]{\cite{#1}}

\begin{document}
	
\renewcommand\linenumberfont{\normalfont}
	
\title[Article Title]{Field-selective criticality in 2D melting revealed by multi-field Lee-Yang zeros}
	
\author[1,2,3]{\fnm{Ling} \sur{Liu}}\email{lingliu@pku.edu.cn}

\author[1,2,3]{\fnm{Fang-Cheng} \sur{Wang}}\email{wfc@pku.edu.cn}

\author*[1,2,3]{\fnm{Qi-Jun} \sur{Ye}}\email{qjye@pku.edu.cn}
\author*[1,2,3]{\fnm{Xin-Zheng} \sur{Li}}\email{xzli@pku.edu.cn}


\affil[1]{\orgdiv{Interdisciplinary Institute of Light-Element Quantum Materials, Research Center for Light-Element Advanced Materials, and Collaborative Innovation Center of Quantum Matter}, \orgname{Peking University}, \orgaddress{\city{Beijing}, \postcode{100871}, \country{P. R. China}}}

\affil[2]{\orgdiv{State Key Laboratory for Artificial Microstructure and Mesoscopic Physics, Frontier Science Center for Nano-optoelectronics and School of Physics}, \orgname{Peking University}, \orgaddress{\city{Beijing}, \postcode{100871}, \country{P. R. China}}}

\affil[3]{\orgname{Peking University Yangtze Delta Institute of Optoelectronics}, \orgaddress{\city{Nantong}, \postcode{226010}, \country{P. R. China}}}

\abstract{
How a two-dimensional solid melts remains unsettled after 60 years of study, as theory, model systems, simulations, and atomic-resolution experiments continue to suggest conflicting scenarios.
The same transition can appear continuous or abrupt depending on how it is observed, where this ambiguity is especially acute in confined water.
Here we study bilayer water under nanoconfinement and ask not only where its phase boundaries lie, but how the system responds to the two fields that drive them: temperature and lateral pressure.
Using Lee-Yang zeros together with enhanced sampling, we find that some phase boundaries are field-selective: the two responses can differ either in continuity itself, or in how strongly they are rounded in finite systems.
This distinction changes the two-step melting picture.
The solid--hexatic transition is field-selective first-order, with the density channel remaining unusually rounded, whereas the hexatic--liquid transition becomes a conventional first-order transition once larger cells reveal a hidden bimodal enthalpy distribution.
This framework organizes the apparent disagreement among confined-water simulations, hard-disk models and AgI experiments by identifying which thermodynamic channel each probe sees.
}


\maketitle


Two-dimensional (2D) melting is a central platform for understanding how dimensionality reshapes phase transitions (PTs)~\cite{nishimori2011elements,Stanley1971}. 
The Kosterlitz--Thouless--Halperin--Nelson--Young (KTHNY) theory provides the canonical picture, in which a 2D solid melts through two continuous PTs separated by an intermediate hexatic phase~\cite{Kosterlitz1973,Nelson1979,Young1979}. 
Two distinct issues are therefore involved in later studies: whether melting proceeds in one step or through a hexatic phase, and whether each transition is continuous or first-order. 
Model and realistic systems already show that the pathway itself can be richer than the KTHNY scenario, with one-step melting occupying a large portion of the phase diagram~\cite{Bernard2011,Kapil2022,Zeng2023}. 
Even when two-step melting occurs, the order of the two transitions remains debated: hard-disk studies favor a continuous--first-order sequence~\cite{Bernard2011}, whereas molecular dynamics (MD) simulations of confined water have suggested a first-order--continuous ``reversed'' one~\cite{Kapil2022,Zeng2023}. 
A recent \textit{in situ} atomic-resolution imaging experiment on 2D AgI supports the hard-disk result~\cite{Bui2025}. 
These apparently conflicting findings show that a unified description of 2D melting requires not only locating the phase boundaries, but also specifying how the transition order is diagnosed in low-dimensional and confined systems. 
One source of ambiguity is that different studies probe different thermodynamic responses. 
Traditional investigations of 2D melting with model or solvable potentials often rely on isotherms to discern the transition order~\cite{Bernard2011,Oscar2021}, while the recent AgI experiment classified melting from density (or volume) evidences~\cite{Bui2025}. 
These observables primarily probe the response to the lateral pressure $p_{\text{L}}$ ($p_{xx}=p_{yy}=p_{\text{L}}$). 
By contrast, MD simulations of molecular systems, including confined water, commonly classify PTs from the evolution of the potential energy ($U$) upon changing temperature~\cite{Zeng2023,Zeng2024,Kapil2022,Stanley2010}. 
This probes the thermal axis, although the variable conjugate to $T$ in the $Np_{\text{L}}T$ ensemble is the enthalpy ($H$), not $U$. 
If the responses to $T$ and $p_{\text{L}}$ are not locked to each other, a single observable can give an incomplete or even misleading classification. 
A field-resolved criterion is therefore needed.
Besides, a related difficulty, that finite-size rounding in averaged observables can make weak first-order transitions appear continuous, should also be addressed.

Lee-Yang theory offers such a criterion by looking for the zeros of the partition function.
The real parts of leading zeros locate the transition, and their imaginary parts measure how the singularity is approached in a finite system~\cite{Lee1952,Yang1952}.
With holding the other fields fixed, the resulting Lee-Yang zeros (LYZs) therefore track the response to the field of interest, rather than mixing all thermodynamic signatures into a single indicator~\cite{Bena2005,Liu2025,Ouyang2024,Rocha2014,Rocha2017}.
This idea has been used to reveal why supercritical Widom lines defined by different response maxima, such as $C_p$ and $K_T$, are diverged~\cite{Ouyang2024}.
With enhanced sampling, these field-specific zeros can now be obtained for realistic molecular systems from well-converged density of states (DOSs), retaining distribution-level information that can be muted in ensemble averages~\cite{Invernizzi2020,Invernizzi2020b,Liu2025}.
Here we use this field-resolved view to revisit the melting of bilayer nanoconfined water.
Specifically, we combine OPES sampling with multi-field Lee-Yang zero analysis to construct the $T$-$p_\text{L}$ phase diagram and to classify the field dependence of its PTs.
The real parts of the Lee-Yang edges locate the boundaries, whereas the relation between the $T$ edges and $p_{\text{L}}$ edges reveals whether thermal and mechanical responses behave conventionally or become decoupled.
Guided by this structure, we analyze the projected DOSs of $H$ and $V$ and show that edge bifurcation occurs when the volume difference between coexisting phases vanishes while the latent heat remains finite.
This produces a field-selective criticality near the maximum of the hexagonal-ice--liquid coexistence line: the response to $p_{\text{L}}$ becomes continuous, whereas the response to $T$ remains discontinuous.
In the two-step melting regime, we find a different manifestation of field selectivity for the solid--hexatic transition, in which unequal finite-size rounding---overlooked in earlier smaller simulations---becomes crucial.
Resolving the hexatic--liquid transition as first-order from observable averages requires a 1024-molecule supercell, whereas the 256-molecule LYZs still give consistent guidance for field-resolved criticality.
Together, these results organize the apparent disagreement with hard-disk simulations and the recent AgI experiment while showing that confined water departs from the original KTHNY continuous--continuous scenario.

\section*{Results}\label{sec2}
\begin{figure*}[htbp]
	\includegraphics[width=1.0\linewidth] {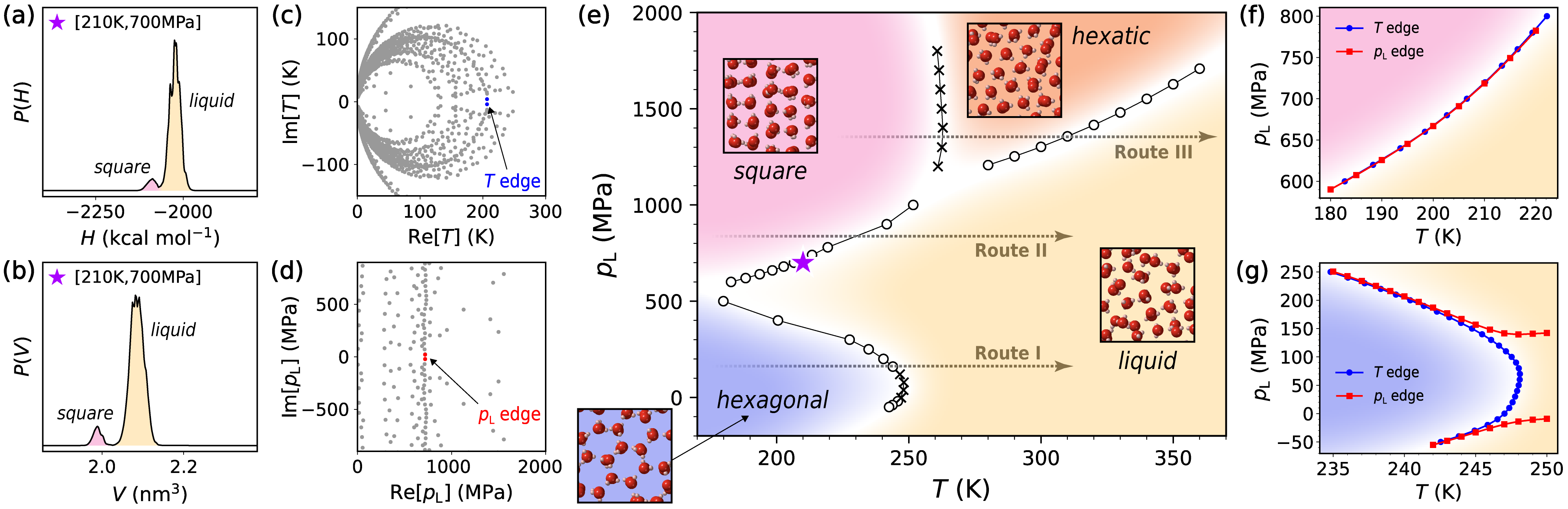}
	\caption{Phase diagram of bilayer nanoconfined water and Lee-Yang zero analysis. a,b, Enthalpy (a) and volume (b) probability distributions sampled at 210 K and 700 MPa. c,d, the corresponding Lee-Yang zeros in the complex $T$ plane at a fixed $p_\text{L}$=700 MPa (c) and those in the complex $p_{\text{L}}$ plane at a fixed $T$=210 K. e, The $T$-$p_\text{L}$ phase diagram of bilayer water. Circular markers indicate conventional first-order transitions where the $T$ edges and $p$ edges coincide. Cross symbols denote regions when a bifurcation occurs between the $T$ edges and $p_{\text{L}}$ edges, and only the $T$ edges are plotted for clarity. Route I (II) highlights the one-step transitions for hexagonal ice--liquid (square ice--liquid) pathways. Route III represents the two-step melting pathway from square ice to liquid. The purple star indicates the ($T, p_\text{L}$) state point used for the sampling analysis in a-d. Insets are the representative snapshots of the liquid, hexatic, square ice and hexagonal ice phases. f,g, The $T$ and $p_{\text{L}}$ edges projected to the real $T$-$p_\text{L}$ plane for the square ice--liquid transitions (f) and for the hexagonal ice--liquid transitions (g).}
	\label{fig:fig1}
\end{figure*}

The main phase diagram for nanoconfined water is constructed from a 256-molecule system modeled by the TIP4P/2005 rigid potential~\cite{Vega2005}. 
Larger 1024-molecule supercells are introduced later only for the finite-size analysis of the two-step melting regime.
These molecules are confined between two hydrophobic walls with a separation of 0.8 nm, a width specifically chosen to accommodate approximately two molecular layers~\cite{Stanley2010}. 
To obtain well-converged DOSs across the relevant basins, we perform OPES simulations~\cite{Invernizzi2020,Invernizzi2020b,Invernizzi2022} within the isothermal-isolateral pressure ensemble ($Np_\text{L}T$). 
By analyzing the $H$ distribution and the $V$ distribution at a given ($T$,$p_\text{L}$) (Fig.~\ref{fig:fig1}a and \ref{fig:fig1}b), we construct the partition functions as functions of complex $T$ (at fixed real $p_\text{L}$) and complex $p_\text{L}$ (at fixed real $T$) to calculate the LYZs (Fig.~\ref{fig:fig1}c and \ref{fig:fig1}d). 
The phase boundaries are then identified by locating the zeros closest to the real axis, hereafter referred to as Lee-Yang edges of $T$ and $p_{\text{L}}$~\cite{Liu2025,Ouyang2024}. All the technical details of calculations can be found in Methods. 
The phase diagram in Fig.~\ref{fig:fig1}e contains four phases: liquid, hexatic, square ice, and hexagonal ice.
One-step melting occupies a large part of the phase diagram, appearing as the hexagonal ice--liquid transition at low pressures (Route I in Fig.~\ref{fig:fig1}e) and the square ice--liquid transition at intermediate pressures (Route II in Fig.~\ref{fig:fig1}e), consistent with earlier simulations of bilayer water~\cite{Stanley2010}.
At higher pressures, melting proceeds in two steps from square ice to hexatic and then to liquid (Route III in Fig.~\ref{fig:fig1}e), a pathway predicted by KTHNY theory and previously reported for monolayer nanoconfined water~\cite{Kapil2022}.
With the boundaries of the phases located, the next question is across each boundary how the different thermodynamic responses behave?
The Lee-Yang edges provide this field-resolved information because each complex field probes the nonanalyticity associated with its conjugate thermodynamic variable.
For conventional PTs, the real parts of the $T$- and $p_{\text{L}}$-edges locate the same boundary, indicating that different conjugate responses change in a coordinated way (Fig.~\ref{fig:fig1}f).
We mark these transitions with circles in Fig.~\ref{fig:fig1}e.
In contrast, we find regions where the $T$- and $p_{\text{L}}$-edges separate (Fig.~\ref{fig:fig1}g), marked by crosses in Fig.~\ref{fig:fig1}e.
Thus, comparing the $T$- and $p_{\text{L}}$-edges exposes field selectivity: the thermodynamic character of a transition can depend on which field is probed, signaled by the edge bifurcation along certain boundaries.
In the following, we shall first analyze Routes I and II to establish the microscopic origin of this signature, and then return to the controversial two-step melting regime along Route III.

\begin{figure*}[htbp]
	\centering
	\includegraphics[width=0.93\linewidth] {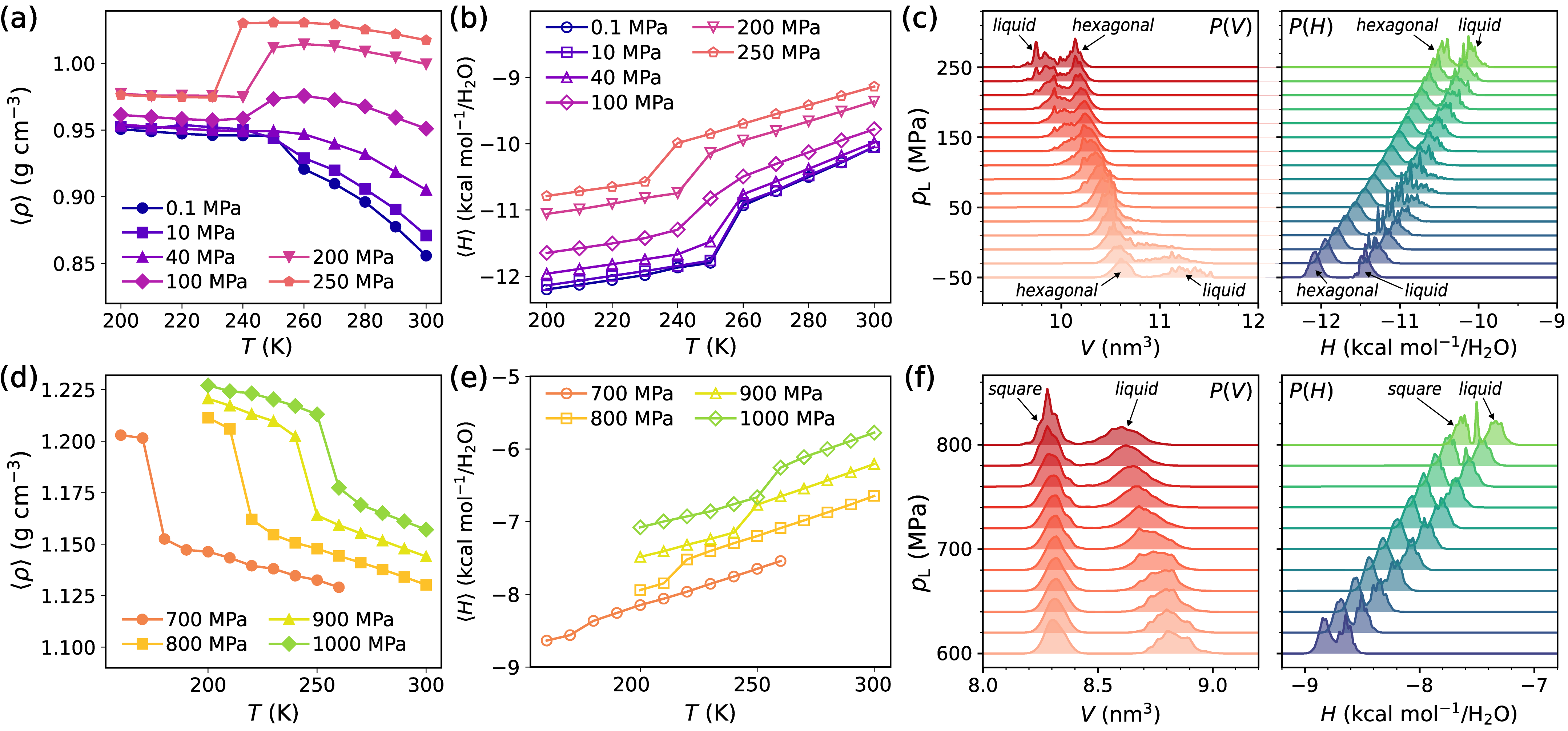}
	\caption{Phase behavior of one-step melting transitions. a,b, Evolution of the ensemble averaged density $\langle \rho \rangle$ (a) and enthalpy $\langle H \rangle$  (b) as a function of $T$ for the hexagonal ice--liquid transition. c, Probability distributions of $V$, $U$ and $H$ along the coexistence line for the hexagonal ice--liquid transition.  $V$ and $U$ represent instantaneous values sampled from simulations, with the enthalpy defined as $H=U+p_\text{L}V$. d,e, Evolution of the ensemble averaged density $\langle \rho \rangle$ (d) and enthalpy $\langle H \rangle$  (e) as a function of $T$ for the square ice--liquid transition. f, Probability distributions of $V$, $U$ and $H$ along the coexistence line for the square ice--liquid transition.}
	\label{fig:fig2}
\end{figure*}
Along Route I for the one-step melting of hexagonal ice, standard $Np_\text{L}T$ MD simulations with gradually decreasing $T$ across a range of $p_\text{L}$ were performed.
The ensemble-averaged densities $\langle \rho \rangle$, which are proportional to $\langle V^{-1} \rangle$, and enthalpies $\langle H \rangle$ as functions of $T$ for $p_\text{L}$ ranging from 0.1~MPa to 250~MPa were plotted in Fig.~\ref{fig:fig2}a and \ref{fig:fig2}b. 
In Fig.~\ref{fig:fig2}a, we see that $\langle \rho \rangle$ evolves from a discontinuous jump at $p_\text{L}=0.1$~MPa to a continuous-like change at $p_\text{L}\sim40$~MPa and back to a discontinuous jump at higher $p_\text{L}$.
In contrast, $\langle H \rangle$ remains discontinuous throughout this pressure range in Fig.~\ref{fig:fig2}b. 
Conventionally, one would expect smooth or sudden changes of the conjugate observables $V$ and $H$ to occur together, reflecting coordinated responses to $p_\text{L}$ and $T$.
The one-step square-ice--liquid transition along Route II provides such an example: $\langle \rho \rangle$ and $\langle H \rangle$ change simultaneously and abruptly (Fig.~\ref{fig:fig2}d and Fig.~\ref{fig:fig2}e).
Therefore, the inconsistent responses for the hexagonal-ice--liquid PT clearly indicate an intricate situation in which relying on a single observable can obscure the order of a PT.
To identify the origin of these different responses, we plot the probability distributions of $V$ and $H$ along the coexistence lines of the two one-step melting processes by reweighting and projecting the DOSs from the OPES simulations. 
The results are shown in Fig.~\ref{fig:fig2}c and Fig.~\ref{fig:fig2}f.
For the hexagonal ice--liquid PT (Fig.~\ref{fig:fig2}c), the $V$ distribution shifts gradually from bimodal to unimodal and again to bimodal as $p_\text{L}$ decreases, whereas the $H$ distributions remain bimodal throughout. 
The overlap of the $V$ peaks, which represents a vanishing volume difference between the two phases, drives the observed continuity in $\langle \rho \rangle$ at $p_\text{L}\sim 40$~MPa in Fig.~\ref{fig:fig2}a.
This phenomenon has a simple thermodynamic origin.
Along a first-order coexistence line, the Clapeyron equation gives

\begin{equation}
	\frac{\partial p}{\partial T} = \frac{\Delta H}{T\Delta V}.
	\label{eq:clapeyron}
\end{equation}
Here, $\Delta H$ and $\Delta V$ denote the latent heat and volume difference between the two coexisting phases, respectively. 
Because $\Delta H$ remains finite, the overlap and subsequent inversion of the $V$ peaks imply $\Delta V\to0$ and then a sign change, so $\partial p/\partial T$ diverges at the rightward-salient $T$ maximum near $p_\text{L}=70$~MPa (Fig.~\ref{fig:fig1}e).
This is not peculiar to confined water: analogous maxima of melting curves, often discussed as reentrant melting, have been observed or predicted in three-dimensional high-pressure systems such as hydrogen and alkali metals~\cite{Deemyad2008,Gregoryanz2005,Hernandez2007,Guillaume2011}.
The same peak structure also explains the field-selective LYZ behavior.
For a conventional transition, the real parts of the $T$- and $p_\text{L}$-edges are expected to track the same phase boundary~\cite{Ouyang2024}, because the singular responses to different thermodynamic fields occur at the same thermodynamic condition.
Near the $T$-maximum of Route I, however, the $V$ peaks overlap while the $H$ peaks remain separated (Fig.~\ref{fig:fig2}c).
This field selectivity appears directly as a bifurcation between the Lee-Yang edges of $T$ and $p_\text{L}$ in Fig.~\ref{fig:fig1}g, and is corroborated by distinct positions for the maxima of the heat capacity and isothermal compressibility (Fig.~S2).
The one-step melting along Route II provide the conventional case: both $H$ and $V$ remain clearly bimodal along this coexistence line (Fig.~\ref{fig:fig2}f), and the $T$ and $p_\text{L}$ edges coincide (Fig.~\ref{fig:fig1}f).
Previous MD simulations identified this transition as continuous~\cite{Stanley2010}, but the thermal response is governed by the distribution of $H$ rather than $U$.
The bimodal $P(H)$ in Fig.~\ref{fig:fig2}f therefore indicates a discontinuous thermal response despite any smoothness inferred from $U$ alone (Fig.~S3).
\begin{figure*}[htbp]
	\centering
	\includegraphics[width=0.9\linewidth] {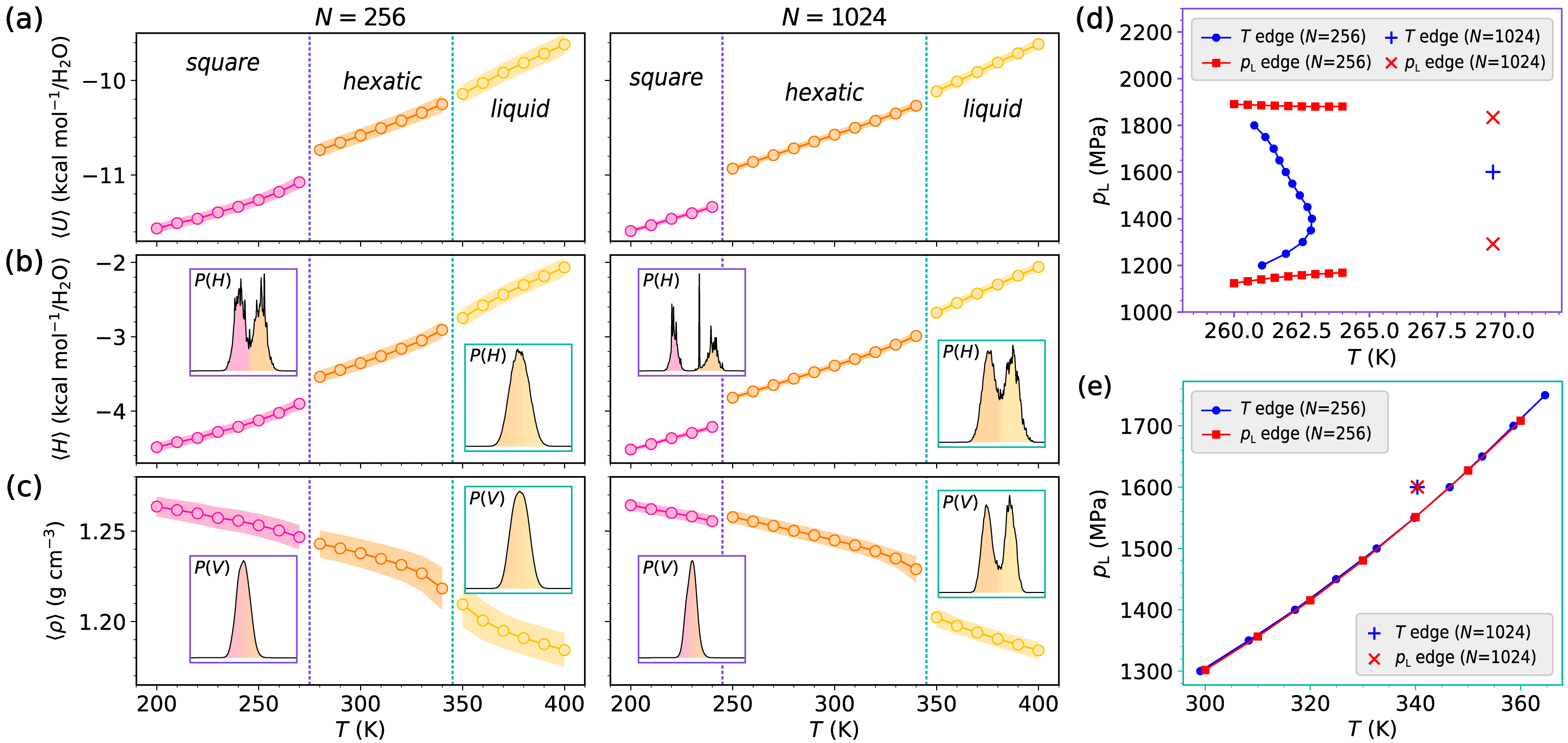}
	\caption{Phase behavior of two-step melting transitions. a,b,c, Evolution of the ensemble-averaged potential energy $\langle U \rangle$ (a), density $\langle \rho \rangle$ (b) and enthalpy $\langle H \rangle$ (c) with decreasing temperature at a fixed $p_\text{L}$ of 1600 MPa for systems containing 256 water molecules ($N$=256, left) and 1024 water molecules ($N$=1024, right).  Data are presented as mean $\pm$ s.d. Insets are the probability distribution of $V$ and $H$ sampled at the equilibrium condition of 1600 MPa. d,e, The $T$ edges and $p_\text{L}$ edges projected onto the real $T$-$p_\text{L}$ plane for the solid--hexatic transition (d) and for the hexatic--liquid transition (e).  }
	\label{fig:fig3}
\end{figure*}

With the one-step Routes I and II clarified, we turn to Route III, where the canonical two-step melting problem is directly at issue.
At $p_\text{L}=1600$ MPa, simulations with 256 water molecules ($N=256$ column in Fig.~\ref{fig:fig3}) reproduce the pattern familiar from previous confined-water MD studies.
$\langle U \rangle$ shows a sharp jump at the solid--hexatic transition followed by a smooth evolution into the liquid (Fig.~\ref{fig:fig3}a), and $\langle H \rangle$ shows a similar apparent sequence (Fig.~\ref{fig:fig3}b).
This behavior was previously interpreted as a first-order solid--hexatic transition followed by a continuous hexatic--liquid transition~\cite{Kapil2022,Zeng2023}.
The LYZ analysis, however, changes how this finite-size result should be read.
For the solid--hexatic transition, the $T$ and $p_{\text{L}}$ edges already bifurcate at $N=256$ (Fig.~\ref{fig:fig3}d), giving the same field-selective signature identified 
along Route I.
It implies that thermal and mechanical responses must be analyzed separately.
The volume peaks remain strongly overlapping across the solid--hexatic transition, making $\langle \rho \rangle$ continuous-like (Fig.~\ref{fig:fig3}c).
At the same size, the hexatic--liquid transition also appears superficially continuous in both $\langle H \rangle$ and $\langle \rho \rangle$, in conflict with hard-disk studies and the AgI experiment~\cite{Bernard2011,Bui2025}.
Because finite-size effects can round bimodal distributions and obscure weak discontinuities in small systems~\cite{Stanley2010}, the LYZ result should be combined with finite-size analysis rather than used as a standalone order classifier.
To make the latent distribution structure more resolvable, we quadruple the system size to a 1024-molecule supercell ($N=1024$ column in Fig.~\ref{fig:fig3}).
For the solid--hexatic transition, $\langle H \rangle$ retains a discontinuous jump, whereas the volume distribution remains nearly unimodal or strongly overlapping, leaving the density response much weaker (Fig.~\ref{fig:fig3}b and Fig.~\ref{fig:fig3}c).
Here, field selectivity manifests as unequal finite-size rounding in the two channels: the enthalpy jump is already clear at $N=256$, whereas the density jump becomes more visible only at $N=1024$.
Accordingly, the solid--hexatic transition is classified as field-selective first-order, even though the density distribution remains strongly overlapping.
The hexatic--liquid transition, by contrast, is not field-selective (Fig.~\ref{fig:fig3}e), so finite-size rounding affects the two responses in a coordinated way.
Both the $V$ and $H$ peaks that are rounded together at $N=256$ become clearly resolved and separated at $N=1024$, establishing the hexatic--liquid transition as first-order.
%

%
Route III composing two first-order transitions therefore differs from the original KTHNY scenario of two continuous ones, but reconciles the apparent disagreement among previous confined-water MD simulations~\cite{Kapil2022,Zeng2023}, hard-disk simulations~\cite{Bernard2011}, and the recent AgI experiment~\cite{Bui2025}.
Previous MD simulations used smaller supercells and mainly followed $U$ rather than $H$, which can obscure the first-order nature of the hexatic--liquid step.
The AgI experiment probed the local density distribution and identified the solid--hexatic step as continuous from its unimodal shape, while we show that a volume distribution can remain unimodal-like even when $\langle\rho\rangle$ already shows a jump.
This caution is consistent with a recent deep-learning study of liquid water, where local density distribution can appear unimodal while learned structural coordinates reveal a bimodal separation of local states~\cite{li2026a}.
This apparent mismatch may reflect the slower approach of the mechanical channel to its thermodynamic-limit behavior, leaving stronger rounding in finite systems.
The same logic applies to hard-disk simulations, where structural correlations were mainly used to identify the solid--hexatic transition as continuous.
The edge bifurcation is less affected by finite-size rounding than raw averages (Fig.~\ref{fig:fig3}d and Fig.~\ref{fig:fig3}e), thus the LYZ analysis provides the guidance needed to decide when separate thermal and mechanical finite-size analyses are required.

\section*{Discussion}\label{sec13}
Beyond resolving the order of each transition in the $T$-$p_{\text{L}}$ phase diagram, the multi-field LYZ analysis shows that a phase boundary need not have a single thermodynamic character.
In bilayer confined water, this appears as a separation of the $T$ and $p_{\text{L}}$ Lee-Yang edges near the maximum of the hexagonal ice--liquid coexistence line and along the solid--hexatic transition (Fig.~\ref{fig:fig1}g and Fig.~\ref{fig:fig3}d).
The separated edges indicate that thermal and mechanical channels can respond differently, and can also be rounded differently by finite size, rendering classification based on a single observable inadequate.
This field selectivity explains why density-, energy-, and enthalpy-based probes can assign different apparent orders to the same melting process.
Such response decoupling is usually discussed above a critical point as distinct Widom lines, but here it occurs on phase-transition boundaries.
Finite-size scaling based on the extensivity of entropy suggests that the edge bifurcation persists toward the thermodynamic limit (Fig.~S4 and S5).
Although field-selective criticality is not necessarily unique to two dimensions, nanoconfinement and reduced dimensionality make it especially important to distinguish thermal and mechanical responses while experiments and simulations often access only a subset of observables.
Since these channels can in principle be tested through response-function measurements, similar field-selective transitions may be identified and help resolving controversies in other confined, anisotropic, or low-dimensional systems.
\section*{Methods}

\subsection*{Computational details}
Enhanced sampling simulations used the OPES method as implemented in the PLUMED plugin~\cite{plumed2014} integrated with the LAMMPS molecular dynamics package~\cite{lammps2022}. 
We mapped the phase diagram using two complementary strategies: 
(i) standalone OPES-explore simulations~\cite{Invernizzi2022} are used to locate individual coexistence points at selected $(T, p_\text{L})$ conditions; 
(ii) OPES-explore was combined with multithermal--multibaric (MTMB) simulations~\cite{Invernizzi2020b,Trizio2024} to sample broad ranges of $T$ and $p_{\text{L}}$ and trace continuous coexistence lines. 
Standalone OPES-explore simulations were run for at least 100 ns, and combined OPES-MTMB simulations for at least 200 ns. 
The bias potential was updated every 0.5 ps. 
We used the translational entropy $s_2$---a widely used descriptor for liquid--ice transitions~\cite{Piaggi2017,Niu2019,Liu2025}---as the collective variable (CV) to drive the phase transitions.
It is defined from the instantaneous oxygen--oxygen radial distribution function $g(r)$ as
\begin{linenomath}
\begin{equation}
	s_2 = -2\pi\rho k_\text{B} \int_{0}^{r_\text{max}} [g(r)\ln g(r)-g(r)+1]r^2 dr,
\end{equation}
\end{linenomath}
where $\rho$ is the system density, $k_\text{B}$ is Boltzmann's constant, and $r_\text{max}=5$ \AA. 
While reweighting the density of states, the first 10 ns of each trajectory was discarded to ensure convergence of the bias potential. 

MD cooling simulations at each $p_{\text{L}}$ were run for 10--150 ns, depending on $T$ and $p_\text{L}$. 
$T$ and $p_{\text{L}}$ were maintained using Nos\'{e}--Hoover thermostat and barostat~\cite{Nose1984,Hoover1985} with relaxation times of 0.1 ps and 1 ps, respectively. 
A time step of 1 fs was employed. 
For the TIP4P/2005 water model, Lennard-Jones (LJ) and Coulomb interactions were truncated at 10 \AA. 
Long-range Coulomb interactions were treated with the slab-adapted particle-particle particle-mesh (PPPM) solver~\cite{Hockney1988,Yeh1999} to an accuracy of $10^{-6}$. 
The SHAKE algorithm constrained the bond lengths and angles of water molecules~\cite{Berendsen1977}.

The system was confined between two smooth hydrophobic walls. 
The interaction between water molecules and each wall was described by a 9-3 LJ potential:
\begin{linenomath}
\begin{equation}
	U(\Delta z)=4\epsilon \left[\left(\frac{\sigma}{\Delta z}\right)^9-\left(\frac{\sigma}{\Delta z}\right)^3\right],
	\label{eq:lj93}
\end{equation}
\end{linenomath}
where $\Delta z$ denotes the perpendicular distance from the oxygen atom of a water molecule to the wall. 
The LJ parameters were set to $\sigma=2.5$ \AA~and $\epsilon=0.2988$ kcal/mol, which are commonly used to represent effective interactions between water and a hydrophobic confining surface~\cite{Lee1994,Stanley1999,Stanley2010,Zeng2022,Zeng2024}. 
The 1024-molecule system was generated by a 2$\times$2 replication of the 256-molecule cell in the directions parallel to the walls, with the wall separation fixed.


The density in this work were calculated using the modified volume $V'=L_x L_y(L_z-2\sigma+r_0)$~\cite{Stanley1999}, where $r_0$ is the LJ radius parameter of the TIP4P/2005 force field, and $L_x$, $L_y$ and $L_z$ denote the simulation-box lengths. 
Further technical details are provided in the corresponding input files within the Source Data repository.

\subsection*{Calculation of Lee-Yang zeros}
Detailed descriptions of Lee-Yang zero calculations from molecular dynamics simulations can be found in Refs.~\onlinecite{Ouyang2024,Liu2025}.
Here we summarize the procedure. 
To compute zeros in the complex temperature plane, we fixed $p_\text{L}$ and treated $T$ as a complex variable $\tilde{T}$. 
The corresponding partition function is
\begin{linenomath}
\begin{equation}\label{eq5}
	Z_p(\tilde{T})= \int \rho(H)e^{-\tilde{\beta} H} \mathrm{d}H,
\end{equation}
\end{linenomath}
where $\rho(H)$ denotes the enthalpy density of states. 
By discretizing the enthalpy with bin size $\Delta H$ (so that $H_k=H_0+k\Delta H$), the integral can be written as a polynomial in $e^{-\tilde{\beta} \Delta H}$:
\begin{linenomath}
\begin{equation}\label{eq7}
	Z_p(\tilde{T})= \Delta H e^{-\tilde{\beta} H_0} \sum_k \rho_k [e^{-\tilde{\beta} \Delta H}]^k.
\end{equation}
\end{linenomath}
At the sampled physical condition ($T,p$), $\rho(H)$ can be obtained from the enthalpy probability distribution $P(H)$ through $\rho(H)=P(H)e^{\beta H}$. 
Substituting this into Eq.~(\ref{eq7}) gives
\begin{linenomath}
\begin{equation}\label{eq8}
	Z_p(\tilde{T}) =  \Delta H e^{-(\tilde{\beta}-\beta) H_0} \sum_k a_k [e^{-(\tilde{\beta}-\beta) \Delta H}]^k.
\end{equation}
\end{linenomath}
Here, $a_k=P_k(H)$. Omitting constant nonzero prefactors reduces the partition function to a polynomial in $y = e^{-(\tilde{\beta}-\beta) \Delta H}$:
\begin{linenomath}
\begin{equation}
	Z_p(\tilde{T})  \sim \sum_k a_k  y^k \sim  \prod_n (y-y^*_n). 
	\label{eq:poly}
\end{equation}
\end{linenomath}

Zeros in the complex pressure plane were computed analogously. 
We fixed $T$ and treated $p$ as a complex variable $\tilde{p}$, giving
\begin{linenomath}
\begin{equation}\label{eq9}
	Z_T(\tilde{p})= \int \rho(V)e^{-\beta \tilde{p} V} \mathrm{d}V,
\end{equation}
\end{linenomath}
where $\rho(V)$ is the volume density of states. 
After discretizing $V$ and substituting the volume probability distribution $P(V)$, we obtain
\begin{linenomath}
\begin{equation}\label{eq10}
	Z_T(\tilde{p}) =  \Delta V e^{-\beta(\tilde{p}-p) V_0} \sum_k b_k [e^{-\beta(\tilde{p}-p) \Delta V}]^k,
\end{equation}
\end{linenomath}
where $b_k=P_k(V)$. 
The partition function then reduces to a polynomial in $x = e^{-\beta(\tilde{p}-p) \Delta V}$:
\begin{linenomath}
\begin{equation}\label{eq11}
	Z_T(\tilde{p})   \sim \sum_k b_k  x^k \sim  \prod_k (x-x^*_k). 
\end{equation}
\end{linenomath}
The roots $y^*_k$ and $x^*_k$ were obtained numerically using a Python script and further converted to $T$- and $p_{\text{L}}$-zeros. 
The enthalpy and volume bin sizes were set to $\Delta H=1$ kcal/mol and $\Delta V=0.01$ nm$^3$, respectively.

To compute Lee-Yang zeros for scaled systems, we used the extensivity of entropy. 
The scaling factor $m$ denotes how many times the reference system size ($m=1$) is amplified. 
Under this scaling, $V\to mV$, $U \to mU$, and the microcanonical entropy satisfies $S(mU,mV)=mS(U,V)$, where $S(U,V)=k_\text{B} \ln \rho(U,V)$ and $\rho(U,V)$ is the number of microstates at given $U$ and $V$. 
The reference density of states $\rho(U,V)$ was obtained from the $m=1$ simulations, leading to
\begin{linenomath}
\begin{equation}
	\rho(mU,mV)=[\rho(U,V)]^m.
	\label{eq:entropy}
\end{equation}
\end{linenomath}
This relation reflects the exponential growth of the density of states with system size. 
Using $\rho(mU,mV)$, we computed the corresponding Lee-Yang zeros and examined their convergence with respect to $1/m$ (Fig. S4 and S5). 
The calculations solving for the zeros at $m>1$ were performed in Mathematica, with the associated scripts provided in the Source Data. 

\backmatter


\bmhead{Acknowledgements}
	This work is supported by the National Natural Science Foundation of China under Grants No. 12550005, No. 12234001, No. 12522410, No. 12474215, and No. 62321004, the National Basic Research Program of China under Grants No. 2021YFA1400500 and No. 2022YFA1403500. We thank the supercomputer center at Peking University for computational resources.

\bibliography{ref}

\end{document}